\documentclass[preprints,article,accept,moreauthors,pdftex]{Definitions/mdpi} 
\usepackage{gensymb}
\usepackage{subfig}

\firstpage{1} 
\pubvolume{xx}
\issuenum{1}
\articlenumber{5}
\pubyear{2019}
\copyrightyear{2019}
\history{}

\Title{MASS-UMAP: Fast and accurate analog ensemble search in weather radar archives}

\Author{Gabriele Franch $^{1,2}$\orcidA, Giuseppe Jurman $^{1}$\orcidB, Luca Coviello$^{1}$\orcidC, Marta Pendesini $^{3}$, Cesare Furlanello $^{1}$\orcidD}

\AuthorNames{Gabriele Franch, Giuseppe Jurman, Luca Coviello, Marta Pendesini, Cesare Furlanello}

\address{%
$^{1}$ \quad Predictive Models for Biomedicine and Environment, Fondazione Bruno Kessler, Trento, Italy franch@fbk.eu (G.F.); jurman@fbk.eu (G.J.); coviello@fbk.eu (L.C.); furlan@fbk.eu (C.F.)\\
$^{2}$ \quad Department of Information Engineering and Computer Science (DISI), University of Trento, Trento, Italy \\
$^{3}$ \quad Meteotrentino, Trento, Italy; marta.pendesini@provincia.tn.it (M.P.)}

\corres{Correspondence: franch@fbk.eu}

\abstract{The use of analogs - similar weather patterns - for weather forecasting and analysis is an established method in meteorology. The most challenging aspect of using this approach in the context of operational radar applications is to be able to perform a fast and accurate search for similar spatiotemporal precipitation patterns in a large archive of historical records. In this context, sequential pairwise search is too slow and computationally expensive. Here we propose an architecture to significantly speed-up spatiotemporal analog retrieval by combining nonlinear geometric dimensionality reduction (UMAP) with the fastest known Euclidean search algorithm for time series (MASS) to find radar analogs in constant time, independently of the desired temporal length to match and the number of extracted analogs. We compare UMAP with Principal component analysis (PCA) and show that UMAP outperforms PCA for spatial MSE analog search with proper settings. Moreover, we show that MASS is 20 times faster than brute force search on the UMAP embeddings space. We test the architecture on real dataset and show that it enables precise and fast operational analog ensemble search through more than 2 years of radar archive in less than 5 seconds on a single workstation.}


\begin{document}

\section{Introduction}
The observation of repeating weather patterns has a long history~\cite{lorenz1969atmospheric}, and the use of analogs has found its way in almost all aspect of meteorology, for the most diverse purposes. Approaches based on analogs have been proposed for the postprocessing of numerical weather predictions~\cite{delle2011kalman}, as a statistical downscaling technique~\cite{zorita1999analog} and for data assimilation in numerical models~\cite{lguensat2017analog, tandeo2015combining}. However, the most prolific use of analogs is by far forecasting: either as a proxy for predictability~\cite{SHAHRIARI2020789}, or as prediction technique itself~\cite{bergen1982long}. On this regard, one of the most used operational methods for analog-based forecasting is Analog Ensemble (AnEn)~\cite{dellemonache2013}, which involves searching and using an ensemble of past analogs to generate new predictions, deterministic or probabilistic~\cite{dellemonache2013}. Ensemble methods have been used for very complex prediction targets, like short-term wind~\cite{alessandrini2015novel} or renewable energy forecasting~\cite{alessandrini2015analog}.
Analogs can be sought to match single locations (pointwise time series) or spatial distributions over an area (spatio-temporal sequences). The quality of the analogs is totally dependent on the dimensionality of the data archive~\cite{van1994searching}: depending on the context, the historical records can spawn from few years to multiple decades, making the analog search procedure a critical component of any operational analog application method. The ideal analog search method should be dependable, predictable, accurate, fast and able to return multiple ranked analogs at the same time. In this paper we present a novel search method for spatio-temporal sequences and show how it meets many of these desirable qualities.

In nowcasting applications (very short-term prediction, between 0 to 6 hours), where the available time for computation is extremely limited, it is often necessary to trade-off analog search accuracy in favor of speed to meet a given computational time threshold. In this regard, one of the most important and complex problems is nowcasting precipitation fields. Analog ensemble approaches based on radar precipitation fields have been proposed for this application~\cite{Panziera2011, SOKOL2017245, altencia2015}: in particular, the AnEn method is especially relevant, since the nowcasting of convective precipitations is extremely challenging to tackle by Numerical Weather Prediction (NWP) methods~\cite{sun2014use}.
AnEn approaches to radar nowcasting use feature extraction~\cite{Panziera2011}, linear dimensionality reduction~\cite{pca2015} or cross correlation~\cite{altencia2015} to summarize and perform an Euclidean search through the radar image archive, in combination with other indicators like mesoscale variables, seasonality and time of the day, to filter the pool of valid sequences. 

In this work we propose a flexible framework that employs a two-step process that can be used to improve the accuracy and dramatically speedup the retrieval of spatiotemporal analog sequences.
Our work combines non-linear geometric dimensionality reduction method based on Uniform Manifold Approximation and Projection~\cite{mcinnes2018umap} (UMAP) with the fastest Euclidean based profile search algorithm (Mueen's Algorithm for Similarity Search~\cite{FastestSimilaritySearch}, MASS), that can match arbitrarily long Euclidean search queries in constant time.
We compare UMAP dimensionality reduction with Principal Component Analysis (PCA)~\cite{jolliffe2011principal} on a original test dataset of almost 10 years of radar data, and show that UMAP finds analogs with smaller Mean Squared Error (MSE) than the one extracted by the PCA based method proposed in~\cite{pca2015}, with proper train configurations parameters. Moreover, we discuss how the MASS search algorithm is 20 times faster than linear Euclidean search and how it can be used to search the reduced UMAP space to find arbitrarily long analog sequences in constant time. Finally we discuss the flexibility of the UMAP-MASS method by showing how other indicators can be easily integrated in the search space to filter analogs and how it is feasible to fine-tune the dimensionality reduction using a custom distance function to project the embeddings.

\section{Materials and Methods} \label{materials}

\subsection{UMAP: Uniform Manifold Approximation and Projection} \label{umap}
Uniform Manifold Approximation and Projection~\cite{mcinnes2018umap} (UMAP) is a recent dimensionality reduction method aiming at reconstructing in a lower dimensional space the geometric structure of the variety where data lie. UMAP is based on algebraic techniques mapping the original manifold into a simplicial complex. Due to its faithfulness in the representation and the low computational burden, UMAP is becoming the reference algorithm for dimensionality reduction in multiple research fields~\cite{becht18dimensionality}.
The dimensionality reduction of UMAP is driven by 4 parameters: $metric$, number of components ($d$), number of neighbors ($n$), minimum distance ($\textrm{mindist})$.

The $metric$ parameter is the distance function used to compare elements in the space of the input data. By default UMAP uses the Euclidean distance function, but any non-negative symmetric function $f\colon\mathbb{R}^2\to\mathbb{R}$ can be a valid UMAP metric.

The number of components $d$ represent the size of the projected space where the transformed data will lie: for example with $d = 4$ every input element will be reduced to a vector of 4 values.

The $n$ parameter controls the number of nearest neighbors used for each point to build the local distance function, taking into account the trade-off between highligthing the details rather than the global picture when rearranging the input data. Using higher values will force UMAP to look at larger neighborhoods to estimate the structure of the data, while using small values will give more weight to local structures present in the data.

The $\textrm{mindist}$ parameter forces a minimum distance between points in the output space. A smaller value will allow UMAP to pack similar elements closer to each other and describe finer topological structure, that can be useful for example for clustering applications, while higher values will preserve the broader structure more.


\subsection{MASS: Mueen's Algorithm for Similarity Search} \label{mass}
One of the most important subroutine in time series analysis is searching for similar patterns to a query string. Any new algorithm or method that is able to speedup this task can potentially open new disruptive applications in any system that deals with time series and historical records. Mueen's Algorithm for Similarity Search~\cite{FastestSimilaritySearch} is probably the most interesting result in the domain of Euclidean based searches in the last few years, where it has spawn new research directions~\cite{matrix1, yeh2018towards, matrix5, matrix8, matrix11}. Recently its application as an analog search function for forecasting time series of renewable energy has been proposed~\cite{ultrafast}. The key idea in MASS is to perform the search in the frequency domain by computing the fast Fourier transform (FFT) on the time series, and replace the loop typically used in similarity matching algorithms with a convolution operation between the query vector and the search archive vector, thus making the search routine independent of the query length. This makes the algorithm free from the curse of dimensionality: matching a long query takes the same time than matching a short one. For this reason the algorithm complexity depends only on the size of the search archive $l_s$ and its complexity is $O(l_s\log{}l_s)$ in the worst case. In the meanwhile, it also produces all the distances from the query to all sub-sequences of the archive, allowing to find all the most similar profile in one single pass. MASS is also parameter-free and can be easily parallelized by splitting the data archive vector in chunks.

\subsection{Meteotrentino Radar Dataset} \label{radar}
The data used for this study was provided by Meteotrentino, the official Civil Protection Weather Forecasting Agency of the Province of Trento, Italy. The agency operates a single polarization Doppler C-Band Radar located on Mt. Macaion (1866 m.a.s.l.), in a very complex orographic environment in the center of the Italian Alps (N46\degree 29’ 18”, E11\degree 12’ 38”).
The radar produces a scan every 5 minutes, for a total of 288 scans per day, and covers an area of 240km of diameter (27,225 sq km).
For this study the product used is the 2D MAX(Z) reflectivity (maximum on the vertical section) at 500mt horizontal resolution. The range of recorded reflectivity values for the product is 0 to 55 dBZ. Every scan is represented as a 480 x 480 floating point matrix over the bounding box coordinates: N47\degree 12’ 39.9168”, S45\degree 46’ 08.6412”, E12\degree 19’45.3593”, W10\degree 07’ 12.7347”.
The radar has been in operation since 2003, at the beginning with different operating modes and frequencies. To guarantee an homogenized product we considered the dates between June 2010 and July 2019, a time span where the radar setup and the quality of the product have been consistent. The total number of radar images in the period is $801,562$, spanning 7 years and 228 days, after accounting for missing data acquisition during the time period.

At least $50\%$ of the data in the archive consists of radar scans with no precipitation that its not interesting for the purpose of analog retrieval, so we decided to remove most of it. This choice has also the effect of reducing the computation time of the analysis, and avoid possible data imbalances, even if UMAP is usually fairly robust against this issue. Since the analog sequence retrieval process expects temporal continuity in the data, we developed a strategy to remove most of the empty data while keeping the temporal discontinuity of the resulting dataset to a usable level. Instead of working on single radar images, we divide the data in chunks of contiguous images by splitting the data by day. Due to missing scans we can obtain more than one chunk of contiguous data for the same day. Chunks longer than 2 hours are kept, the rest is discarded, so each chunk contains at least 25 and at most 288 contiguous scans. Finally we thresholded all chunks with no or few precipitation events: the sum of all pixel values of all images of each chunk is computed, and all chunks with an average pixel value $< 0.5$ dBZ are discarded. After this cleaning steps, the dataset resulted in $342,598$ radar images, corresponding to 3 years and 95 days of precipitation data.
For the reason stated above, and to avoid temporal overlapping, we split the dataset temporally between search space and verification: the data from the years 2010-2016 was used as archive (search space), and the years 2017-2019 as verification (query data). The final result is $l_s = 220,050$ (2 years and 34 days) images for search space and $l_v = 122,548$ (1 year and 61 days) for verification.

\begin{figure}[H]
    \centering
    \includegraphics[height=10 cm]{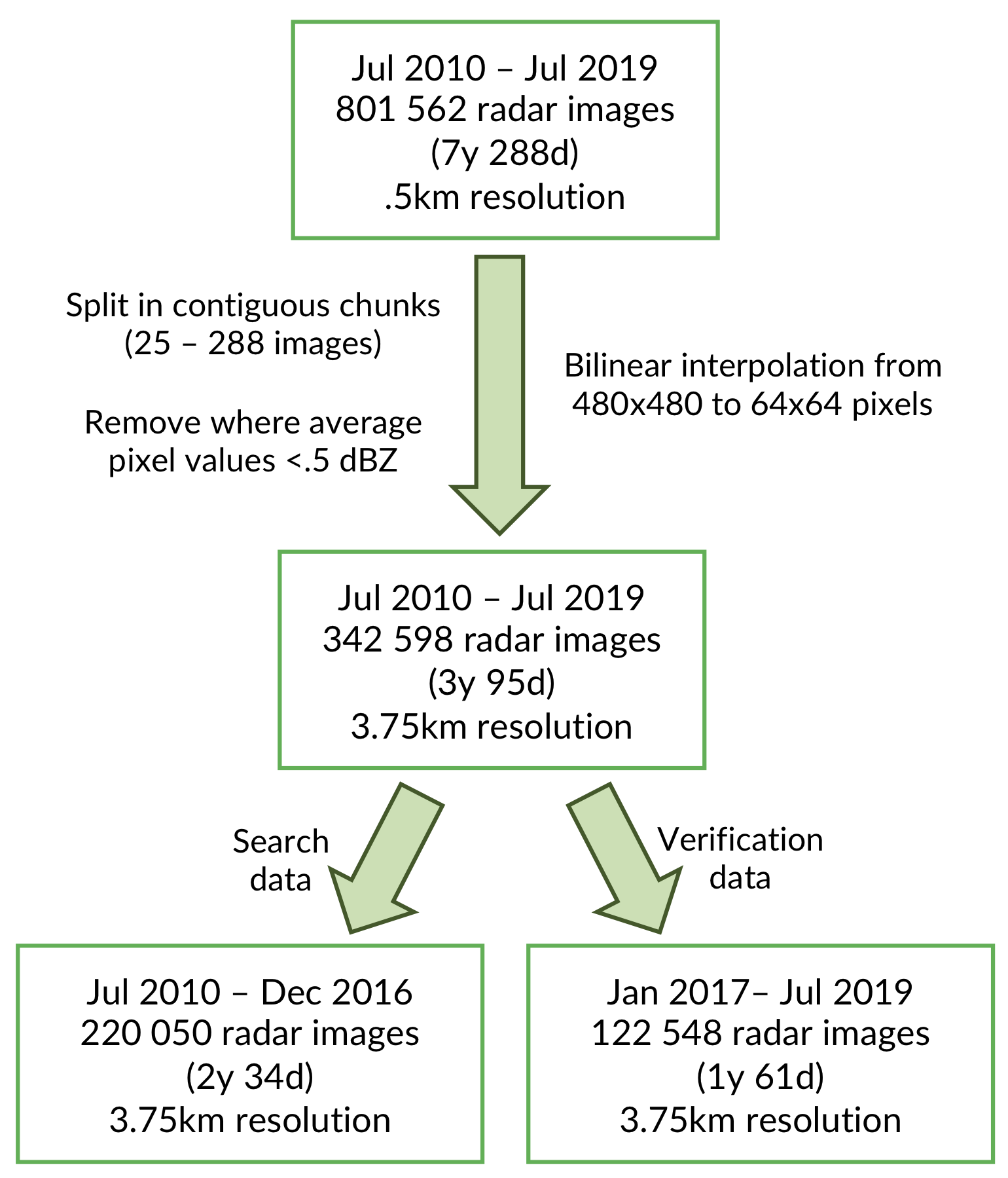}
    \caption{Data pre-processing pipeline. The whole dataset is first filtered to remove data chunks that do not contain a interesting amount of signal. A bilinear interpolation filter is applied to the images to reduce the resolution from 480x480 to 64x64 pixels. The transformed dataset is then split into search and verification sets.}
    \label{fig:dataprep}
\end{figure}

A simple bilinear resize was applied to all the selected data to obtain 64 x 64 pixel images, corresponding to a resolution of 3.75 x 3.75 km. This was chosen for similar reasons to the ones described in~\cite{pca2015}: to reduce the computational time of the experiments and extend the range of tested parameters combinations, to remove small scale variability, and in our case to also alleviate the noise and scatter present int the MAX(Z) product. Figure \ref{fig:dataprep} summarizes the data pre-processing pipeline.

\subsection{MASS-UMAP workflow} \label{massumap}
The workflow of our method follows two phases: dimensionality reduction phase and search phase.

In the first phase we used UMAP to reduce the dimensionality of each image in the radar archive to a vector of length $d$ (embedding), such that $d \ll p$ where $p$ is the number of pixels in one image. UMAP will learn a transformation that maps closer together in a geometric space the input images that are closer according to a specified $metric$. In our experiment we chose the Euclidean distance as indicator $metric$ to compare the images (more details for this choice are discussed in Section~\ref{umapt}), but more specialized metrics are possible. The embeddings are generated for both the search space data and the query data.

\begin{figure}[H]
    \centering
    \includegraphics[height=5.4 cm]{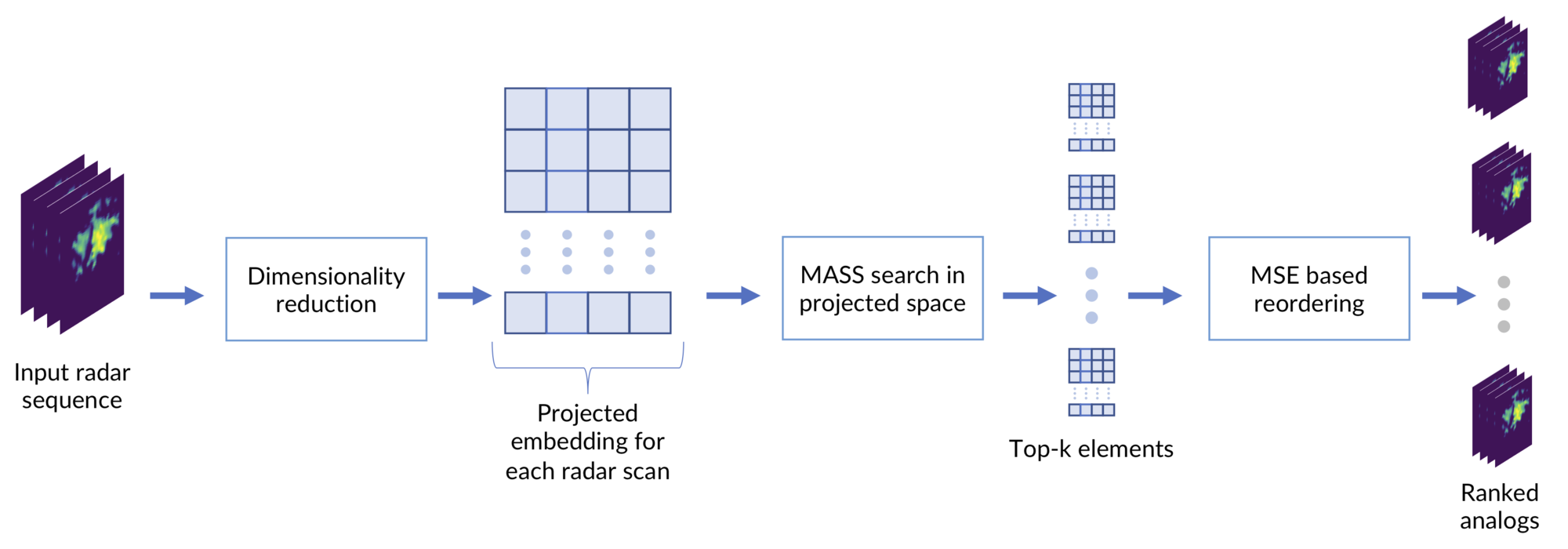}
    \caption{The MASS-UMAP workflow.}
    \label{fig:workflow}
\end{figure}

The second phase uses the MASS algorithm to search through the embedding search space and find the closest matches for the query data. To this aim, the embeddings of all the images in the search space are concatenated together following their natural time order. The result is a vector of length $l_r = d * l_s$. The embeddings of the queries are concatenated in the same fashion, generating vectors of size $q = d * t$ for each query, where $t$ is the time length of the search query (the number of images in the sequence). The MASS algorithm is used to compare the query vector with the search space vector and extract the indexes of a desired number $k$ of closest Euclidean profiles. The use of the Euclidean search is possible because the embeddings are projected by UMAP into a geometric space. As last step, the image sequences corresponding to the indexes of the top $k$ profiles are compared and reordered by computing the MSE respect to the query image sequence, generating the final analog ranking. The desired number of final analogs can be selected by slicing the MSE reordered $k$ analog vector to the final desired size of top-$a$ analogs, with $a \leq k$.
There are two reasons why performing a partial reorder of the top $k$ result before selecting the final analogs is useful: the first is that, even if the dimensionality reduction method works ideally on all cases and always returns the same top $k$ items that an MSE search would, there is hardly any guarantee that those will be returned in the same order.
The second reason is that MASS is able to perform such a fast search because it compares the vector profiles in the frequency domain: while the computed rankings for the profiles are exact, it may match a sequence with a very similar profile of the query vector but with a constant shift on all coordinates. This is indeed a rare occasion, but the partial MSE reordering is useful to move those spurious matches at the bottom of the ranking. So, while we want to avoid computing MSE between the query and all the archive, we can afford a small configurable number of MSE comparisons that can greatly improve the quality of the final ranking. The schema of all steps of the workflow are summarized in Figure \ref{fig:workflow}. 

In an ideal setting, we want the embedding length $d$ to be as small as possible (to reduce the memory requirements and the search time), and to be able to keep $k$ as close as possible to $a$, to minimize the computation needed by the partial reorder step.
For these reasons, we tested the performance of our method with different values of $k$ and $d$ with regard to the ability to return analogs in the same order that an MSE search would.

\subsection{Evaluation Framework}
To better highlight the contribution of the two phases to the overall result, we divided our evaluation in two parts. In the first part (\ref{umapt}) we assessed the performance of different UMAP configurations on single images using two different metrics (\ref{canberra}, \ref{jaccard}) and in comparison with PCA. In the second part (\ref{sequsearch}) we used MASS in conjunction with the best performing dimensionality reduction models, to test the ability of the overall solution in finding analogs of different length ($t$), using the error computed respect to sequences found by MSE search as ground truth. Computational and memory requirements were also considered.

\subsection{Evaluation part I: dimensionality reduction training and verification} \label{umapt}
In Section~\ref{umap} we discussed the 4 parameters driving the UMAP dimensionality reduction algorithm. For the purpose of this study we were especially interested in testing two of them: number of components $d$ (that corresponds to the embedding length) and number of neighbors $n$. Default settings were used for the two other parameters (\textrm{metric}=Euclidean and $\textrm{mindist}=0.1$).
The rationale for this choice is that for the aim of analog retrieval we were not interested in the absolute distance values, but our objective is to keep the same ranking distance between the elements in the original and the embedded space. This means that any distance function that preserves ranking with regard to MSE can be used, such as the default Euclidean distance used by UMAP. The same holds true for the minimum distance between the points where the projected data will lie. This allowed us to concentrate our effort on the remaining parameters where we chose to setup a grid of 6 values for $d = [2, 5, 10, 15, 20, 100]$ and 6 values of $n = [5, 10, 50, 100, 200, 1000]$ for the model optimization.

Using as input the whole set of search data, we fit 36 UMAP transformations with different parameters, given by the cartesian product of 6 choices of $d$ and $n$. We then proceed to produce the embeddings of all the UMAPs for both the search data and the query data.

To compare the ability of UMAP to return good analogs we took the daunting task of computing the MSE distance matrix between all the images in the archive vs all the images in the query set, thus generating a matrix of $l_s * l_v = 220,050 * 122,548 = 26,966,687,400=\approx 2.7\times 10^9$ distances. This matrix allowed us to create an extensive and accurate verification setup of the ability of UMAP to rank and find the same analogs compared to MSE, considering different thresholds of top $k$ elements.

\begin{figure}[H]
    \centering
    \includegraphics[height=10 cm]{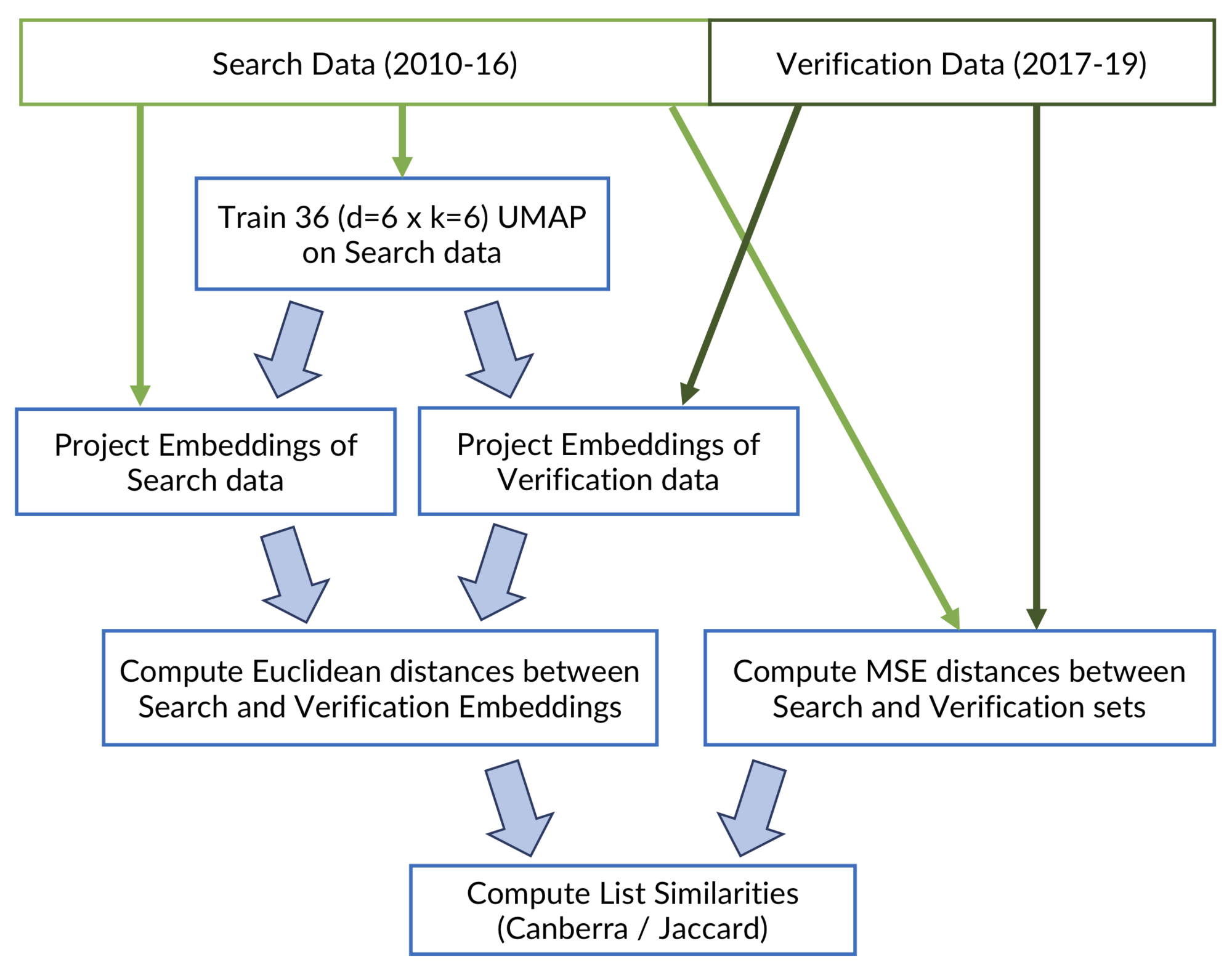}
    \caption{Workflow of the model development for the UMAP training and verification. The same workflow is used for training and verification of the PCA, that it is used as a comparison method.}
    \label{fig:umapwf}
\end{figure}

Figure \ref{fig:umapwf} illustrates the workflow of the UMAP model training and verification. 

As a baseline comparison method we considered the embedding space generated with Principal Component Analysis (PCA)\cite{jolliffe2011principal}: its use as dimensionality reduction technique for radar analog search has been proposed in\cite{pca2015}. Like UMAP, the PCA embedding maps the original space to a geometric space, so we can use Euclidean calculation to compute the distances. All the steps described for UMAP were applied also for PCA with only two notable differences: the first one is that we had to train only one configuration since PCA is non-parametric, the second is that we had to apply more normalization steps to our data archive before computing the PCA, because of the variance maximization used to find the principal components is extremely sensible to unbalanced values.
For this step, we followed the same workflow described in \cite{pca2015}: we applied Box-Cox transformation~\cite{boxcox2012} adding a 0.01 offset to the rainfall rate of each radar scan, and centered each element by removing the corresponding mean before computing the PCA.

We tested all the trained dimensionality reduction configurations methods by introducing two evaluation metrics that helped us to measure the characteristics of the ranking results returned by UMAP/PCA against the ideal rankings found via MSE. The first metric is a weighted ranking correctness measure (\ref{canberra}), the second measures the proportion of correct elements found in the nearest top $k$ (\ref{jaccard}).

\subsubsection{Stability of ranked lists} \label{canberra}
The Canberra stability indicator~\cite{jurman2008algebraic} $I_\textrm{Ca}(\mathcal{L})$ is a group-theoretical measure for assessing the similarity of a set $\mathcal{L}$ of ranked lists of $n$ shared items. The indicator is based on the Canberra distance~\cite{lance1966computer}, a weighted version of the L1 norm whose main features is to penalize more the differences occurring in the top part of the ranked list rather than those occurring at lists' bottom. The indicator is normalized by the expectation $E$ of the Canberra distance on the whole permutation group $S_n$ of cardinality $n!$, so that $0\leq I_\textrm{Ca}(\mathcal{L}) \leq \max_{\rho,\sigma\in S_n}\{\textrm{Ca}(\rho,\sigma)\}\approx 1.42$, with $I_\textrm{Ca}(\mathcal{L})\approx 0$ denotes a set $\mathcal{L}$ of very similar lists, while $I_\textrm{Ca}(\mathcal{L})\approx 1$ indicate that $\mathcal{L}$ is a randomly ranked set of lists~\cite{jurman2009canberra}. By using the locator parameter $k$~\cite{jurman2009canberra}, the same measure can be restricted to evaluate the similarity of the top-$k$ sublists of $\mathcal{L}$ including only the highest ranked items. We used this measure to evaluate how well the top $k$ embedding elements are ranked compared to the MSE ranking.

\subsubsection{Jaccard distance} \label{jaccard}
The Jaccard distance is a dissimilarity measure between two sets. The Jaccard distance ($J_d$) is the complementary of the Jaccard index ($J$)~\cite{jaccard1912distribution}, and it is defined as:

$$ J_d = 1 - J(A,B) = 1 - {{|A \cap B|}\over{|A \cup B|}} $$

where $A$ and $B$ are two set of elements and $\cap$/$\cup$ are the intersection/union operators.

Intuitively this distance helps to understand what is the proportion of items that are present in both sets, normalized by the total number of distinct elements. A value of $J_d = 0$ is obtained between two identical sets, while a value of $J_d = 1$ corresponds to two disjoint sets. We used this measure to evaluate how many of the top $k$ elements found by searching the embedding space corresponded to elements returned by the top $k$ MSE search. The comparison is also useful to assess and compare the performances between PCA and UMAP for analog retrieval. As with the Canberra stability indicator, the Jaccard distance is used to evaluate the similarity of the top-$k$ sublists of $\mathcal{L}$.

\subsection{Evaluation part II: Sequence search evaluation} \label{sequsearch}
In the second part of the evaluation we assessed the complete workflow: we combined UMAP and MASS to test the retrieval of analog sequences of different lengths ($t$) and different number of top-$k$ sequences. For this part we evaluated the solution by comparing the straight cumulative MSE error between the sequences found by MASS-UMAP and MASS-PCA and the sequences retrieved by MSE. We found this comparison to be a more faithful representation of the performances of the overall solution, than using the two metrics introduced in part I. Indeed, since the testing occurs with sequences of different lengths, the total number of possible matches available between the query data and the archive is different for every value of $t$: this makes the interpretation of the metrics much less straightforward.
We also benchmark the use of computing resources required both theoretically and experimentally. Wall execution times are reported when available and discussed, along with some projected scenarios. 

\section{Results} \label{results}

\subsection{Exploration of UMAP embeddings} \label{umap-viz}
\begin{figure}[H]
    \centering
    \subfloat[search set]{{\includegraphics[width=7cm]{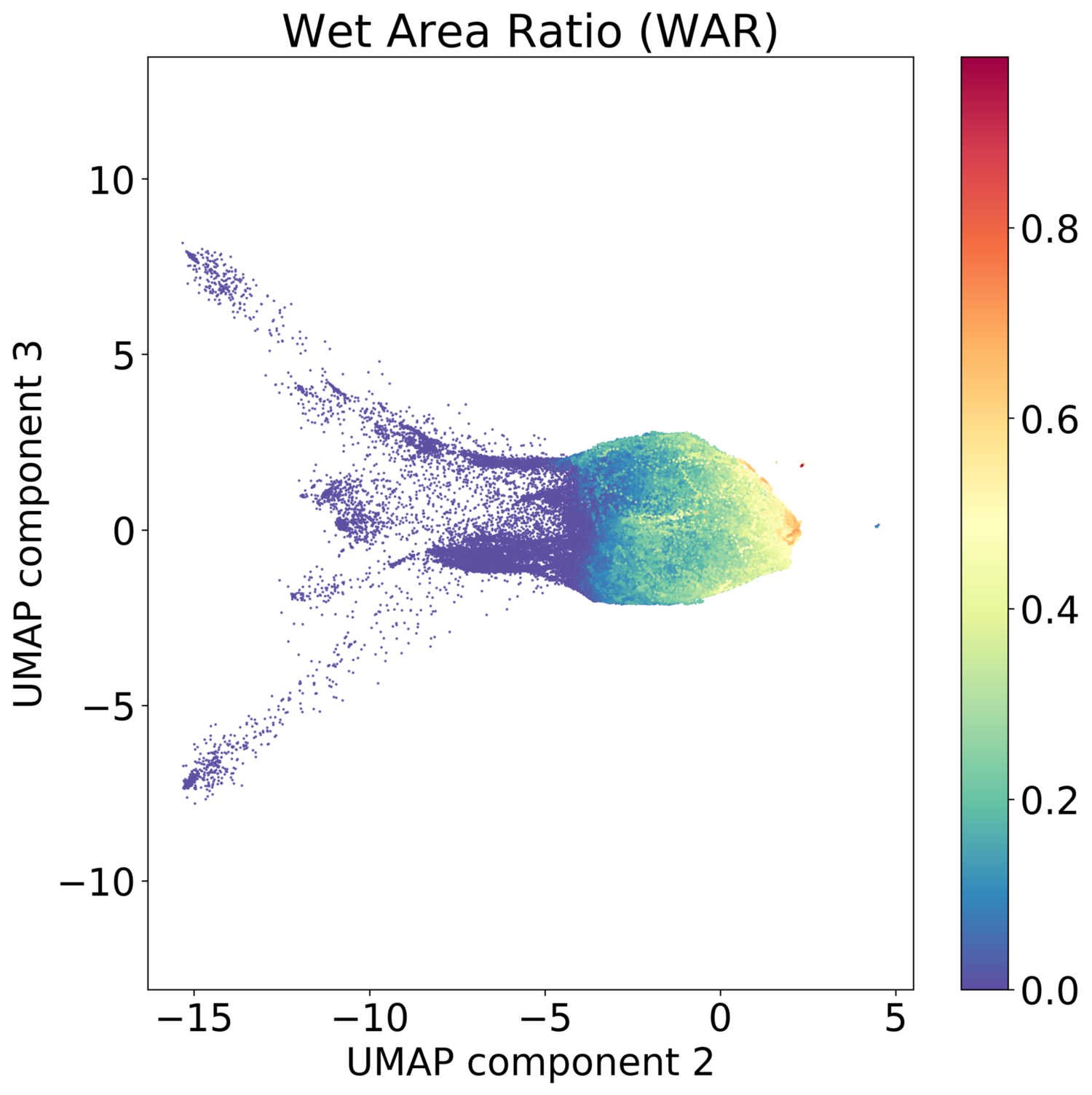} }\label{fig:umap_emb_viz_train}}%
    \qquad
    \subfloat[verification set]{{\includegraphics[width=7cm]{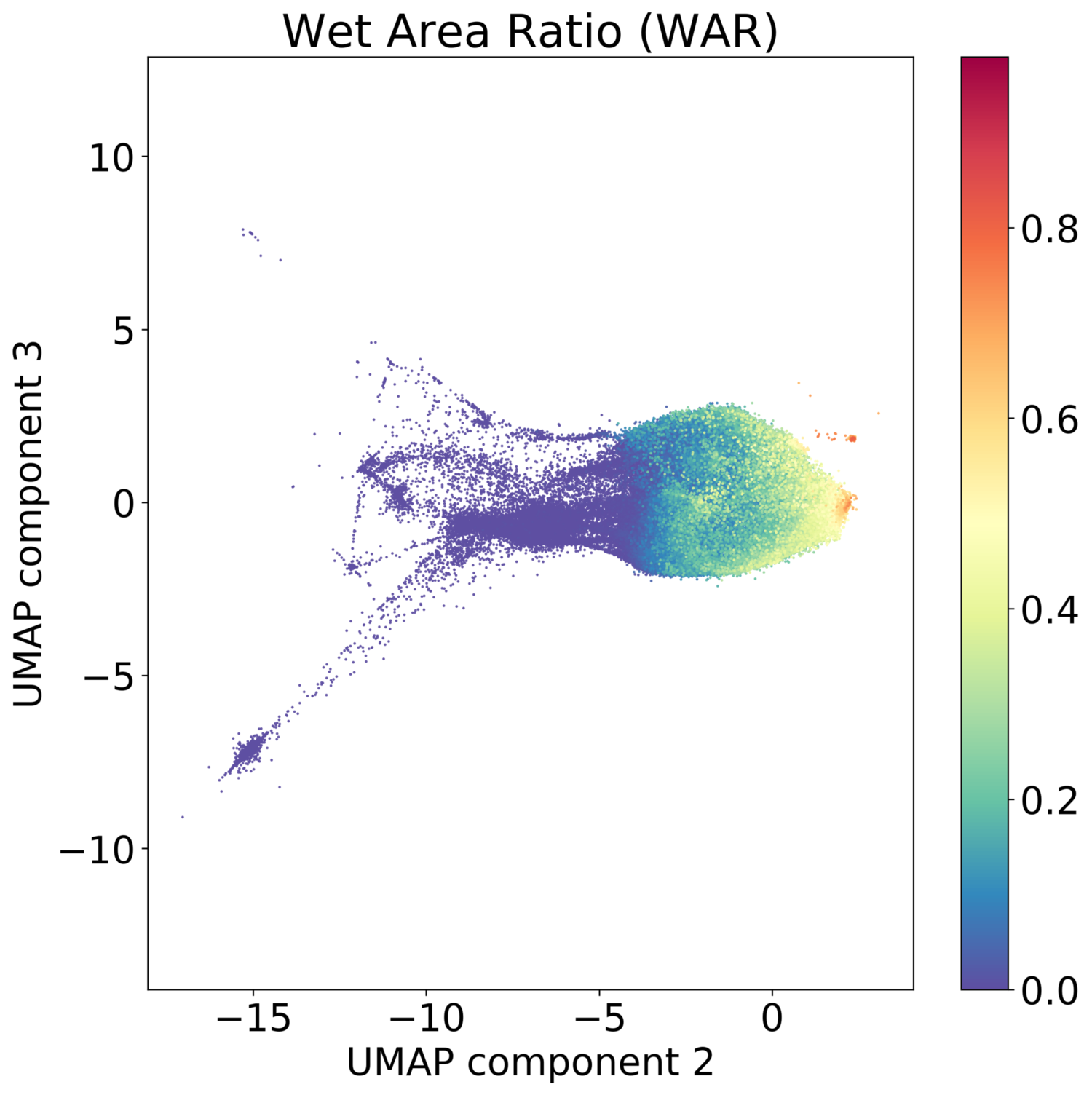} }\label{fig:umap_emb_viz_val}}%
    \caption{UMAP embedding visualization of the second and third components for search space \protect\subref{fig:umap_emb_viz_train} and for verification space \protect\subref{fig:umap_emb_viz_val}. The embeddings are colored by Wet Area Ratio (WAR).}%
    \label{fig:umap_emb_viz}%
\end{figure}

We explored some of the manifolds generated by UMAP projections and plot the resulting embeddings for the search and the verification sets (Figure \ref{fig:umap_emb_viz}). The hyper-parameters for this model are the number of components $d = 5$ and the number of neighbor $n = 200$. If we consider for example the second and third components, the visualization of the two embeddings belonging to the search set (on which the UMAP is constructed) and the verification set are quite similar, where the embedding points are colored by the Wet Area Ratio (WAR), defined as the percentage of pixels with a rain rate higher than $0.1$ mm/h. The stability analysis shows that UMAP is able to project the space maintaining the general distances between radar scans with different rain rates. Notably, the two sets are composed of radar scans collected from 2010 to 2016 for the search set, and from 2017 to 2019 for the verification set. UMAP generalizes well across the two years, and that it is applicable to scans coming from future time windows without the need for retrain.

\subsection{Evaluation part I: dimensionality reduction} \label{single-step-search}
The evaluation of the dimensionality reduction step was implemented as explained in \ref{umapt}. 36 UMAP configurations were fitted on the search data, and the embedding of both search and verification data were generated. The UMAP embeddings were used to compute on the fly the ranking distances for all the verification images and compared with the MSE ranking to compute the cardinality of the set intersections and the Canberra indicator for a number of top-$k$ results. The average and standard deviation for Jaccard and Canberra distances on all validation images were then computed for all possible permutations of $k$ (limits), $d$ (components) and $n$ (neighbors). The final number of computed results is given by the cross product of the configuration space build with the following parameters:
\begin{itemize}[leftmargin=*,labelsep=5.8mm]
\item	limits: $k = 8$ with configurations $K = [5, 10, 15, 20, 50, 100, 200, 500]$
\item	components: $d = 6$ with configurations $D = [2, 5, 10, 15, 20, 100]$
\item	neighbors: $n = 6$ with configurations $N = [5, 10, 50, 100, 200, 1000]$
\end{itemize}
The total number of parameters tested for the UMAP projection is $k * d * n = 8 * 6 * 6 = 288$. Conversely, for PCA the size of the tested configuration space was $k * d = 8 * 6 = 48$, where $D$ is mapped to limit the number of principal components used by the PCA decomposition to the same number of components of UMAP.

\begin{figure}[H]
    \centering
    \includegraphics[height=3.3 cm]{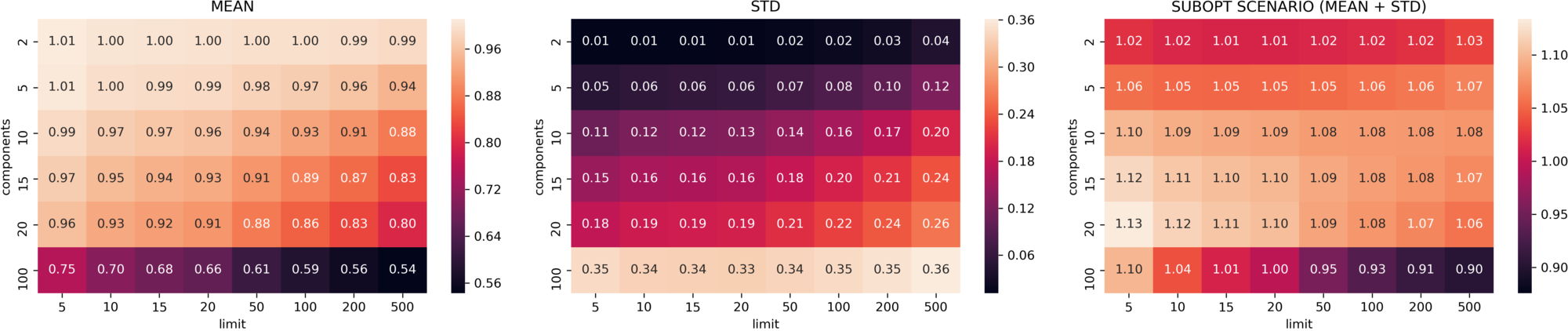}
    \caption{Canberra stability indicator results for PCA with  different values of limit $k$ and components $d$ (darker/lower is better). Mean, standard deviation and the suboptimal scenario given by sum of mean and standard deviation are reported.}
    \label{fig:pcac}
\end{figure}

\begin{figure}[H]
    \centering
    \includegraphics[height=3.3 cm]{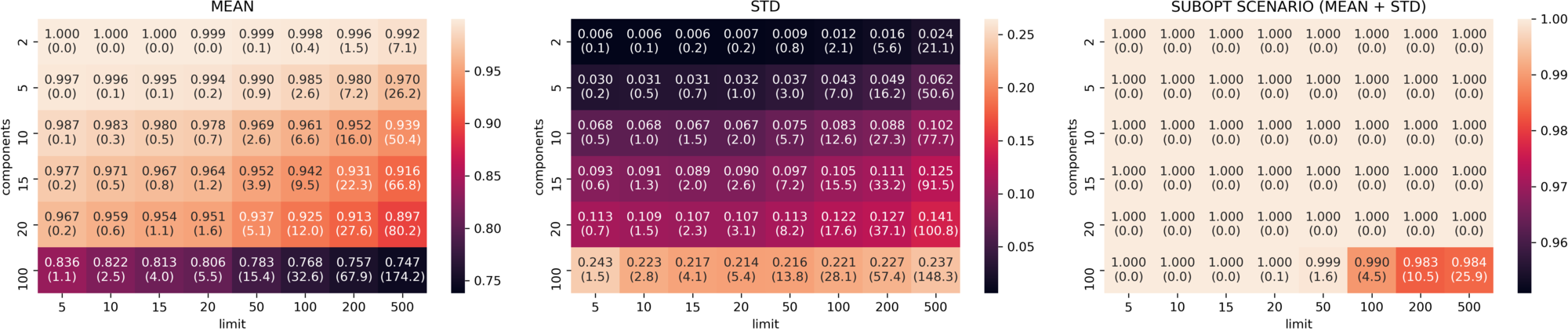}
    \caption{Jaccard values for PCA with different values of limit $k$ and components $d$ (darker/lower is better). The number in parentheses is the cardinality of the intersections between the top-$k$ PCA list and the top $k$ MSE list. Mean, standard deviation and the "worst case scenario" given by the sum of mean and standard deviation are reported.}
    \label{fig:pcaj}
\end{figure}

The results of the 48 configurations tested on PCA are reported in Fig. \ref{fig:pcac} for Canberra stability index and in Fig.~\ref{fig:pcaj} for Jaccard. For each configuration we group together the means, the standard deviations, and the suboptimal scenario, namely the sum  of mean and standard deviation, describing the retrieval performance of the dimensionality reduction with suboptimal results.

Reduction by PCA is consistent and predictable: it shows systematic linear improvements in both Canberra and Jaccard metrics by adding more components and extending the size of top-$k$ considered elements. On the other hand, PCA needs to use at least 20 components and 500 top-$k$ elements to start showing consistently good ranking results (an average of 80 elements in common with the top 500 MSE elements).

\begin{figure}[H]
    \centering
    \includegraphics[height=3.3 cm]{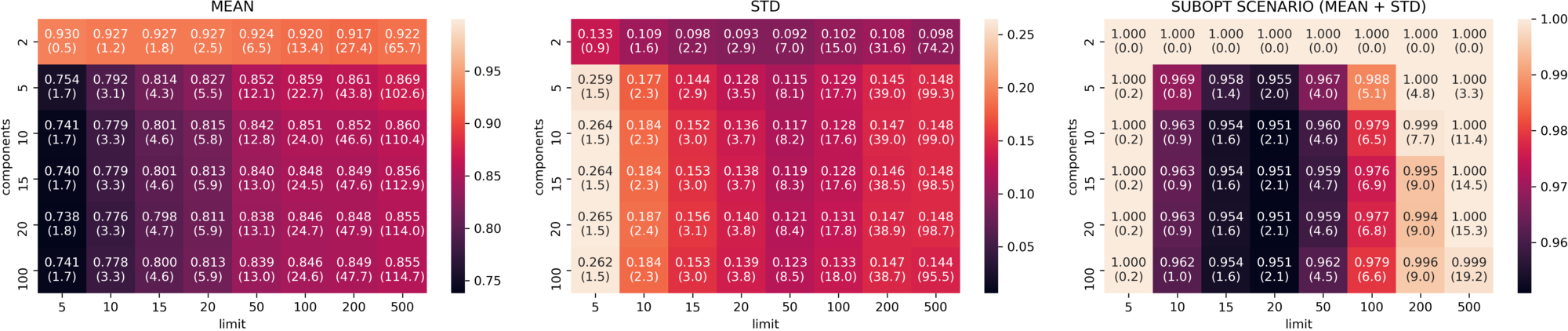}
    \caption{Jaccard results for UMAP models trained with neighbors $n = 5$.}
    \label{fig:umapj5}
\end{figure}

\begin{figure}[H]
    \centering
    \includegraphics[height=3.3 cm]{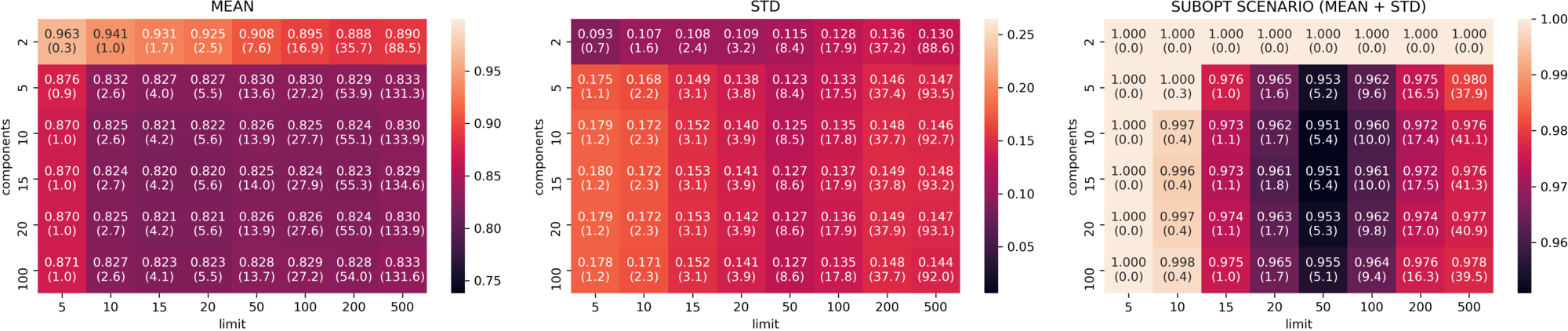}
    \caption{Jaccard results for UMAP models trained with neighbors $n = 10$.}
    \label{fig:umapj10}
\end{figure}

\begin{figure}[H]
    \centering
    \includegraphics[height=3.3 cm]{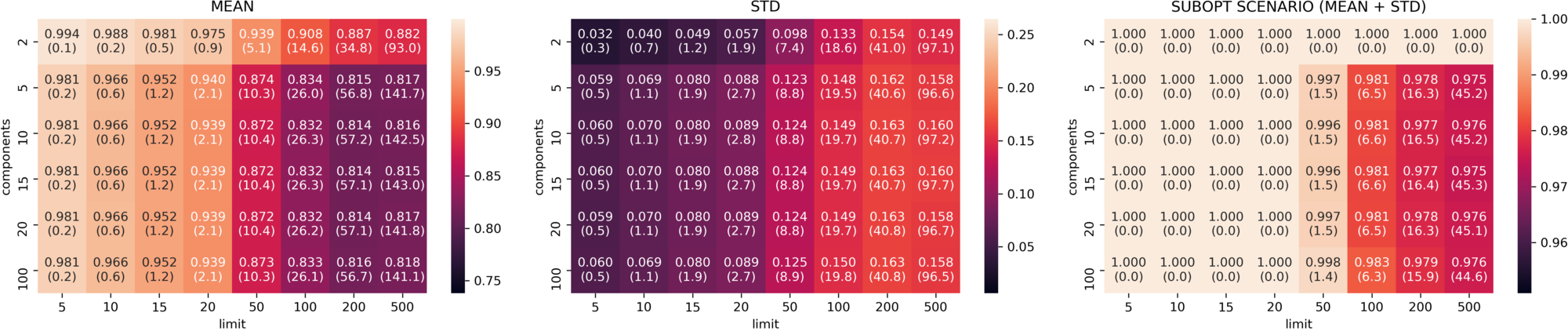}
    \caption{Jaccard results for UMAP models trained with neighbors $n = 50$.}
    \label{fig:umapj50}
\end{figure}

\begin{figure}[H]
    \centering
    \includegraphics[height=3.3 cm]{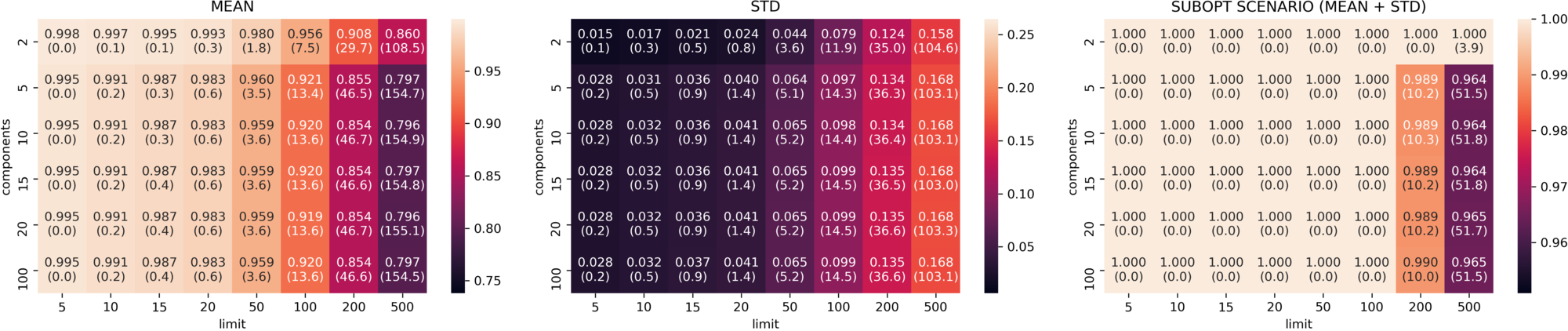}
    \caption{Jaccard results for UMAP models trained with neighbors $n = 200$.}
    \label{fig:umapj200}
\end{figure}

\begin{figure}[H]
    \centering
    \includegraphics[height=3.3 cm]{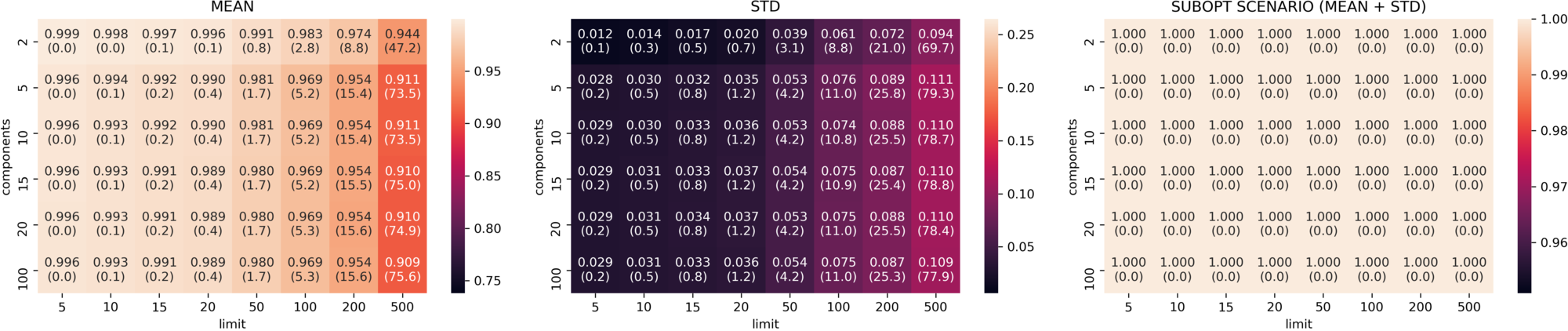}
    \caption{Jaccard results for UMAP models trained with neighbors $n = 1000$.}
    \label{fig:umapj1000}
\end{figure}

In Figs.~\ref{fig:umapj5}, \ref{fig:umapj10}, \ref{fig:umapj50}, \ref{fig:umapj200} and \ref{fig:umapj1000} the analysis of search reduction by Jaccard distance for different values of $n$ of UMAP are reported. The first observation is that UMAP does not follow a linear trend: the algorithm improves dramatically (~40\%) between 2 and 5 components, to then subsequently plateau. Going from 5 to 100 dimensions makes hardly any difference in the ability to find better analogs, and this behavior is consistent even with different values of $n$. Thus, we conjecture that this saturation in the dimensionality is dataset dependent, and that UMAP has already maximized it's ability to describe the data manifold using 5 components. On the other hand, choosing a different value for $n$ drastically changes the performance of UMAP with regard to the choice of $k$. The two values appear to be positively correlated: to train a transformation that finds a consistent number of good analogs in the top-$k$, we need to set $n$ around the value of $k$ (usually a step lower). Given the consistently good results that UMAP showed using just 5 dimensions and the positive correlation between $n$ and $k$ we choose the configuration with $d = 5$, $n = 200$ and $k = 500$ as a benchmark for the second part of our evaluation. In Appendix \ref{extrafigures}, we report the plots for all the remaining configurations not included in this section.

\subsection{Evaluation part II: spatiotemporal analog search performance} \label{multi-step-search}

\subsubsection{Analog quality} \label{analogqual}
\begin{figure}[H]
    \centering
    \includegraphics[height=4.5 cm]{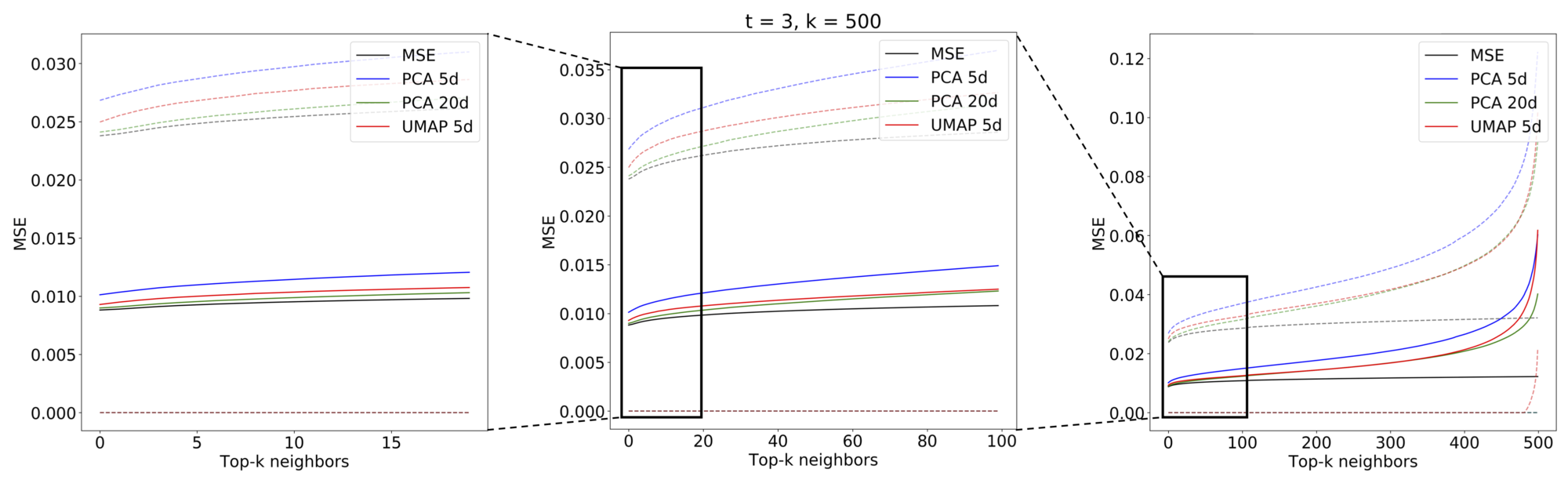}
    \caption{Mean MSE values for analog sequences of $t=3$ obtained with PCA ($c=5$ and $c=20$ components), UMAP ($c=5$ components) and MSE search in original space. Dotted lines represent the standard deviaton of the MSE.}
    \label{fig:umap_pca_t3}
\end{figure}

\begin{figure}[H]
    \centering
    \includegraphics[height=4.5 cm]{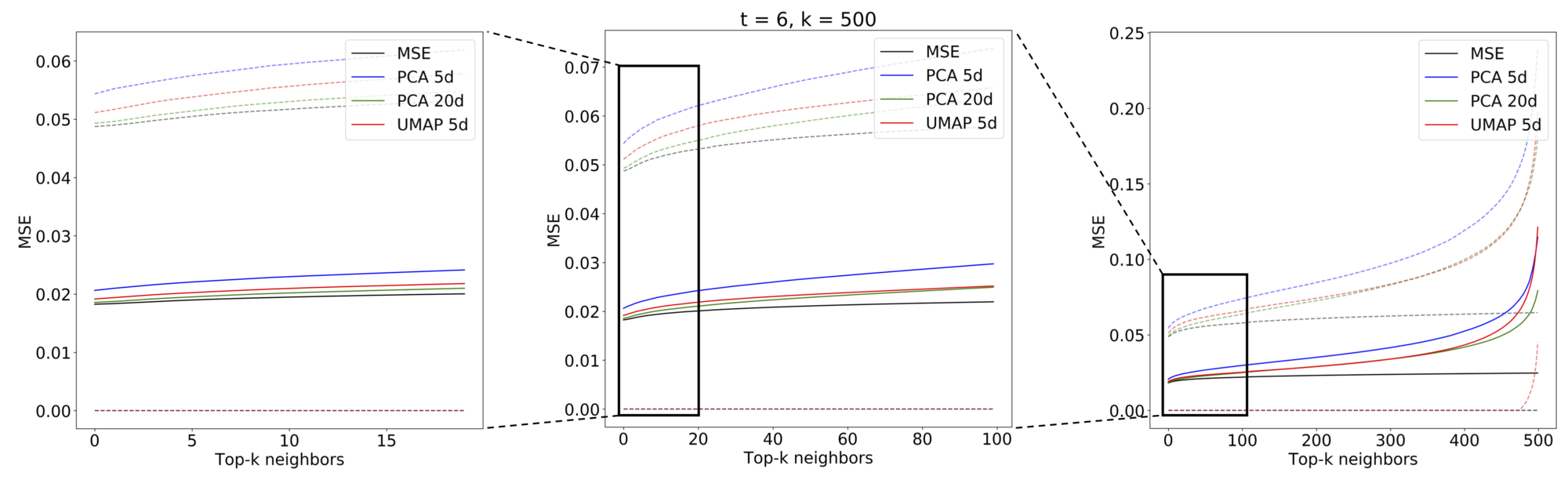}
    \caption{Mean MSE values for analog sequences of $t=6$ obtained with PCA ($c=5$ and $c=20$ components), UMAP ($c=5$ components) and MSE search in original space. Dotted lines represent the standard deviaton of the MSE.}    
    \label{fig:umap_pca_t6}
\end{figure}

\begin{figure}[H]
    \centering
    \includegraphics[height=4.5 cm]{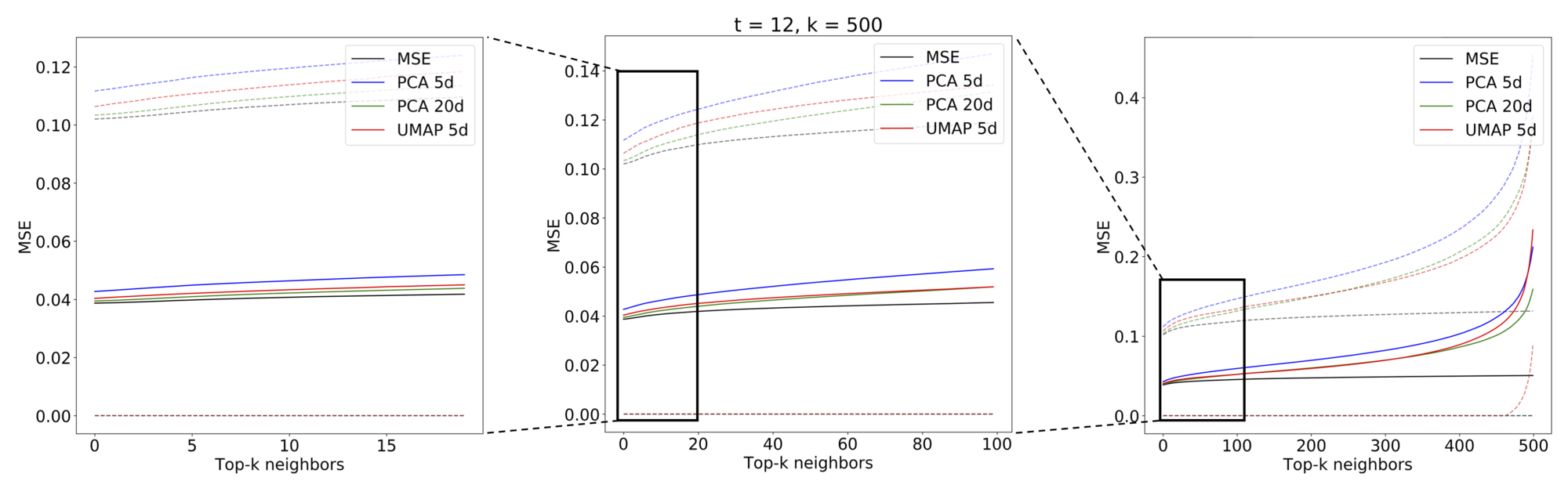}
    \caption{Mean MSE values for analog sequences of $t=12$ obtained with PCA ($c=5$ and $c=20$ components), UMAP ($c=5$ components) and MSE search in original space. Dotted lines represent the standard deviaton of the MSE.}    
    \label{fig:umap_pca_t12}
\end{figure}

\begin{figure}[H]
    \centering
    \includegraphics[height=4.5 cm]{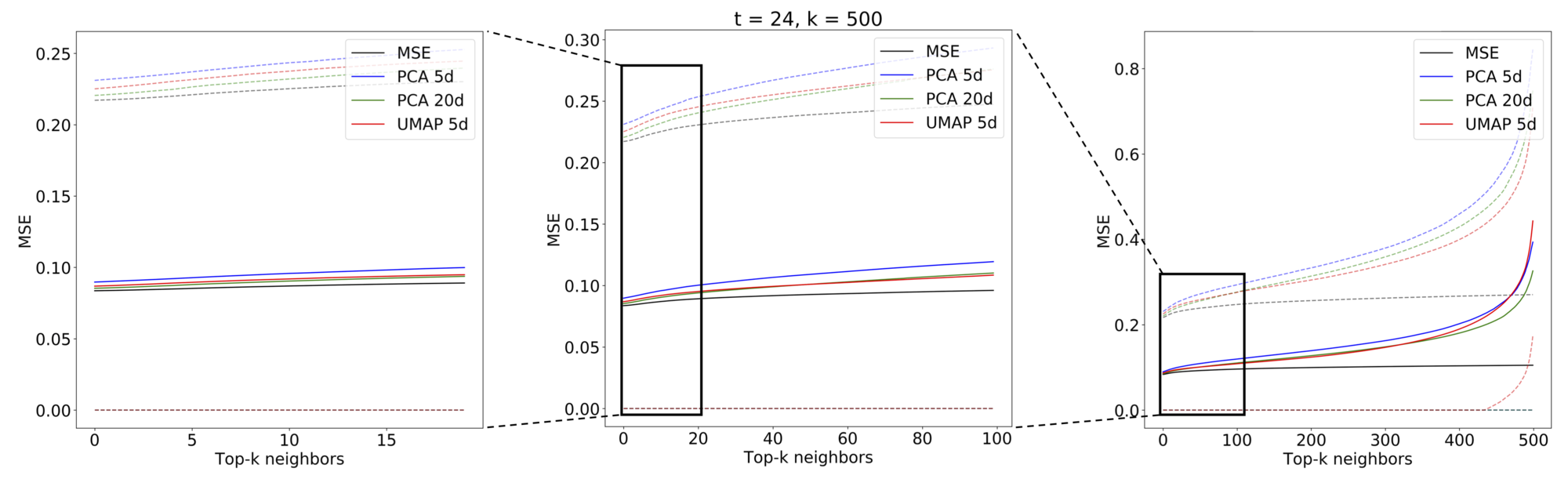}
    \caption{Mean MSE values for analog sequences of $t=24$ obtained with PCA ($c=5$ and $c=20$ components), UMAP ($c=5$ components) and MSE search in original space. Dotted lines represent the standard deviaton of the MSE.}    
    \label{fig:umap_pca_t24}
\end{figure}

The evaluation of the spatiotemporal analog search performance was implemented as explained in Section~\ref{sequsearch}. We used the combination of UMAP and MASS (MASS-UMAP) to find analogs for sequences of $t = 4$ different lengths. We tested values $T = [3, 6, 12, 24]$ corresponding to sequences of 15, 30, 60 and 120 minutes respectively. For comparability, we used the same number of sequences with the same start times for all values of $T$. The sequences were chosen from the query set, starting from the first index and leaving whenever possible 100 images of gap between the beginning of the next sequence: this avoided sequence overlapping, also such gap was sufficiently long to guarantee complete spatiotemporal de-correlation between the chosen sequences. The total number of extracted sequences after such processing was 1226.
We compared the best UMAP configuration ($d = 5$, $n = 200$) with respect to the sequences found by MSE and the sequences found by PCA with 5 and 20 components.
The figures \ref{fig:umap_pca_t3}, \ref{fig:umap_pca_t6}, \ref{fig:umap_pca_t12}, \ref{fig:umap_pca_t24} show the mean MSE, of the models with different values of $T$ and a MSE reorder on the top $k = 500$ elements.

\begin{figure}[H]
    \centering
    \includegraphics[height=3.5 cm]{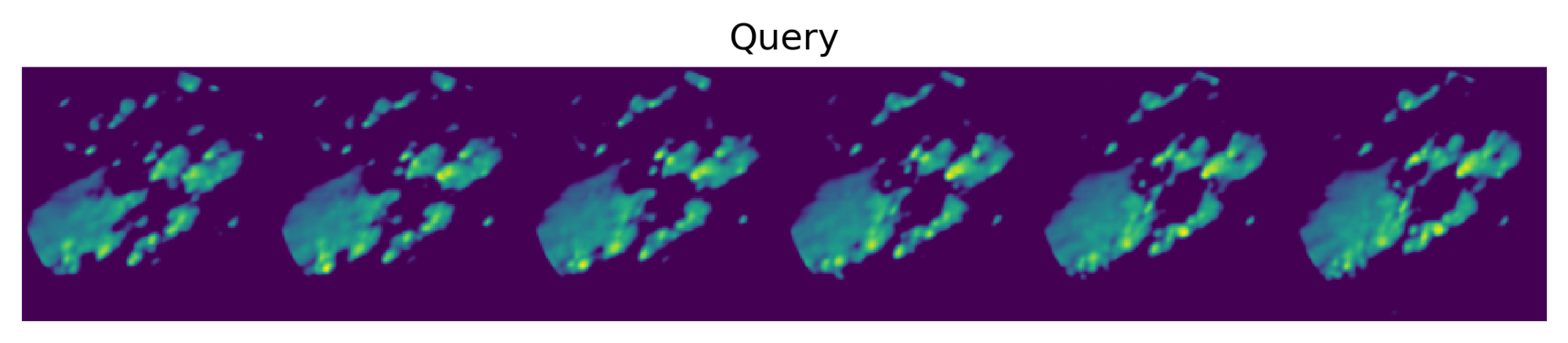}
    \caption{Sample query sequence of $t=6$ radar scans sampled from the verification set.}
    \label{fig:query}
\end{figure}

\begin{figure}[H]
    \centering
    \includegraphics[height=7 cm]{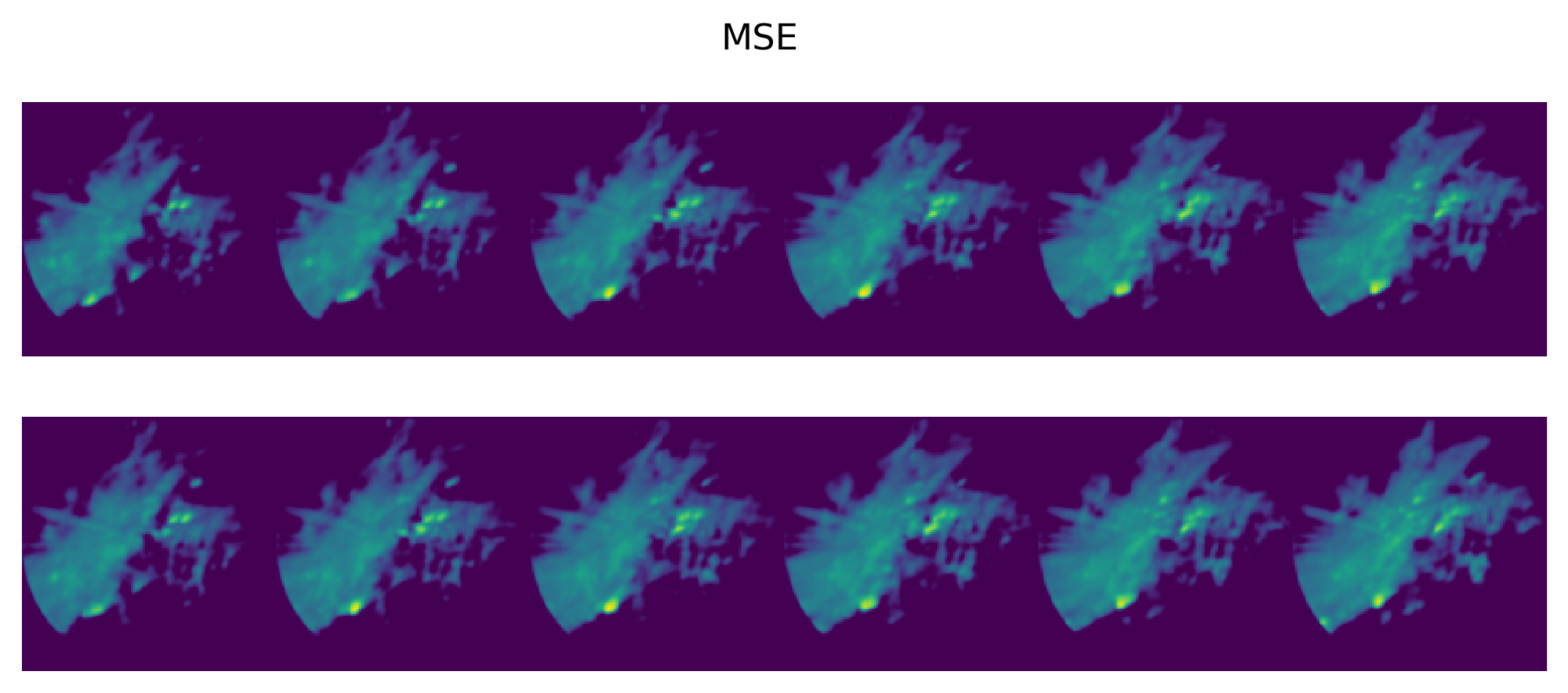}
    \caption{Top-2 most similar sequences as per MSE computed on the original radar scans.}
    \label{fig:mse}
\end{figure}

\begin{figure}[H]
    \centering
    \includegraphics[height=7 cm]{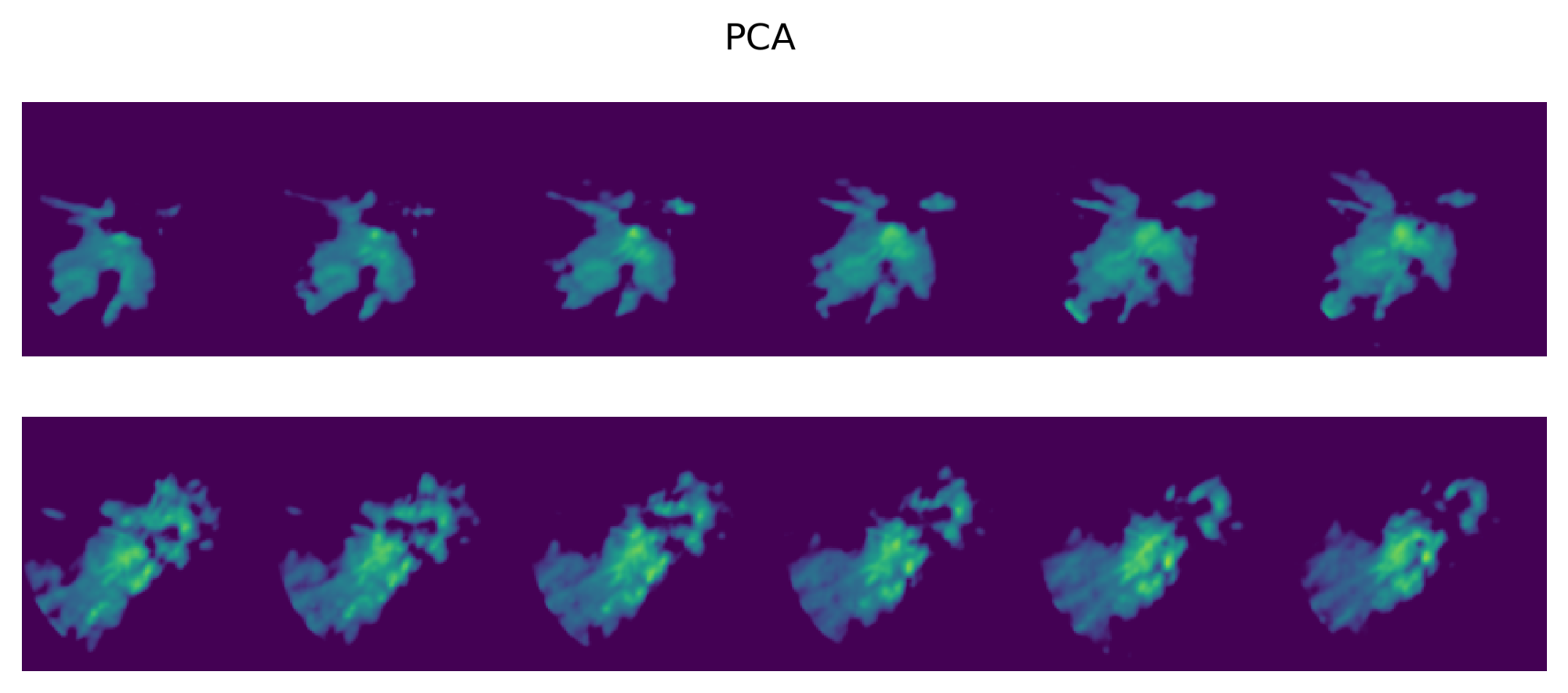}
    \caption{Top-2 most similar sequences resulting from the search on PCA embeddings with MASS.}
    \label{fig:pca}
\end{figure}

\begin{figure}[H]
    \centering
    \includegraphics[height=7 cm]{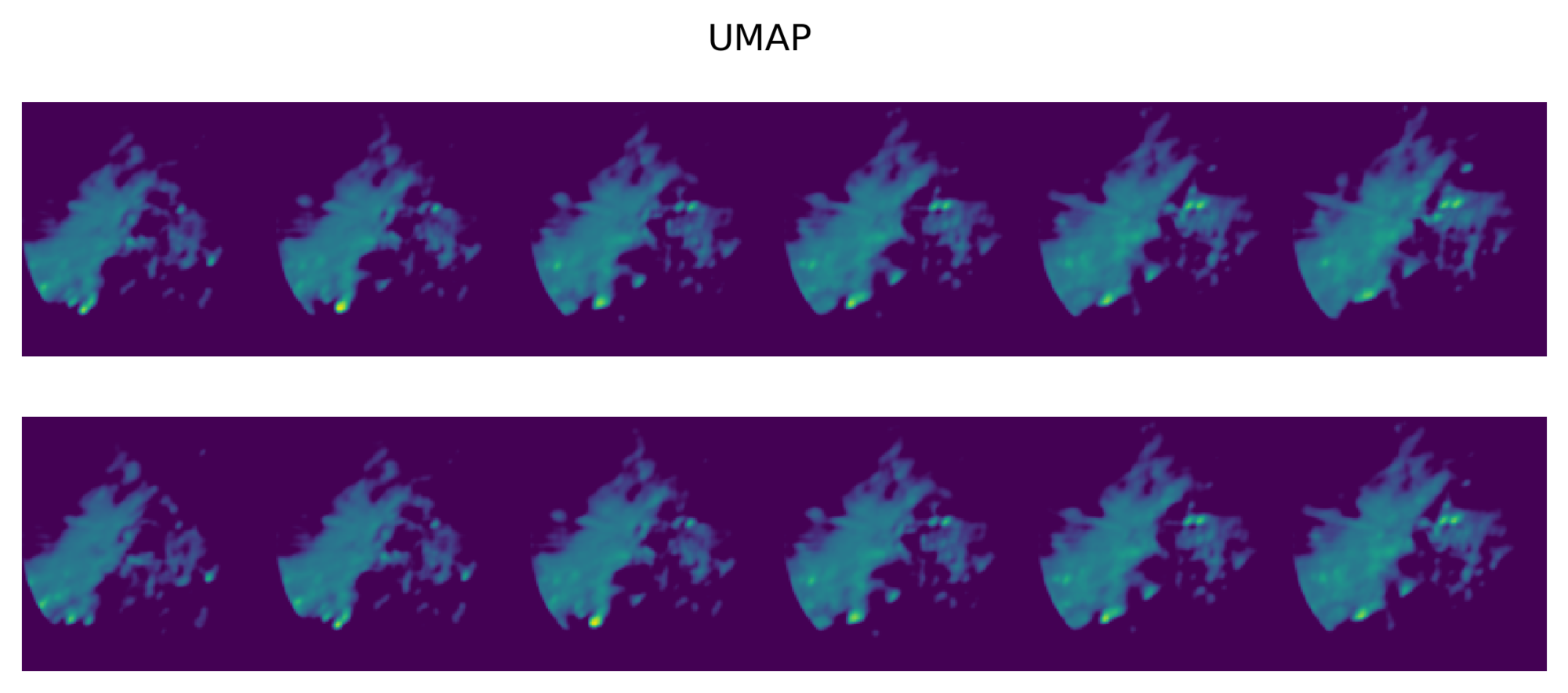}
    \caption{Top-2 most similar sequences resulting from the search on UMAP embeddings with MASS.}
    \label{fig:umap}
\end{figure}

The figures show the plot of the average MSE error between the query sequences and the top $k=500$ sequences for each $i$-th analog in the rank.  
The results are consistent with the performance on single images: the 5 component UMAP consistently outperforms 5 components PCA on all $T$ by a wide margin. The UMAP results are actually on par with 20 components PCA, where UMAP accounts for slightly higher MSE error for $t = 3 $ and $t = 6$ an lower for $t = 12$ and $t = 24$. 
Results of the top-2 most similar sequences found given a query sample (Figure \ref{fig:query}) of $t=6$ radar scans are shown for the 3 compared methods: using MSE on the original scans (Figure \ref{fig:mse}) and using MASS and MSE-based reordering on the top-500 closer embeddings on PCA (Figure \ref{fig:pca}) and on UMAP (Figure \ref{fig:umap}) embeddings. It is visually clear that UMAP is able to resemble more closely the sequences that are closer for MSE and with regard to PCA.

\subsubsection{Execution times and memory requirements}
We tested the execution times of MASS-UMAP on our dataset first by benchmarking each component of the method separately and then by executing the whole workflow end to end. Tab.~\ref{tab:benchmark} shows the result of our benchmarking. The UMAP configuration chosen for the test is the same used in \ref{analogqual} with parameters $d = 5$, $n = 200$ and $k = 500$.
\begin{table}[H]
    \centering
    \begin{tabular}{lrrrr}
        \toprule
        Sequence Length &                3  &                6  &                 12 &                 24 \\
        \midrule
        1) UMAP Transform &  194 ms $\pm$ 6.72 ms &  303 ms $\pm$ 8.87 ms &   451 ms $\pm$ 11.3 ms &   745 ms $\pm$ 15.5 ms \\
        2) MASS search    &  1.01 s $\pm$ 9.11 ms &  1.05 s $\pm$ 13.4 ms &   1.12 s $\pm$ 23.1 ms &     1.31 s $\pm$ 25 ms \\
        3) top-k MSE reorder    &  11.1 ms $\pm$ .12 ms &  43.6 ms $\pm$ .72 ms &  86.4 ms $\pm$ 1.27 ms &   172 ms $\pm$ 1.11 ms \\
        MASS-UMAP (1+2+3) &  1.22 s $\pm$ 15.6 ms &  1.37 s $\pm$ 23.0 ms &   1.66 s $\pm$ 35.7 ms &  2.23 s $\pm$ 35.67 ms \\
        MASS-UMAP end-to-end &  1.18 s $\pm$ 22.5 ms &  1.37 s $\pm$ 48.4 ms &   1.65 s $\pm$ 82.9 ms &    2.3 s $\pm$ 11.9 ms \\
        linear MSE search &   9.59 s $\pm$ 1.08 s &    20.4 s $\pm$ 1.6 s &    39.5 s $\pm$ 3.74 s &  1min 24s $\pm$ 1.02 s \\
        MASS-UMAP speedup &         \textbf{8.1x} &        \textbf{14.9x} &         \textbf{23.9x} &         \textbf{36.5x} \\
        \bottomrule
    \end{tabular}
    \caption{MASS-UMAP execution times comparison with linear MSE search}
    \label{tab:benchmark}
\end{table}

All reported tests were executed ten times; confidence intervals are reported. The test platform used was a non-burstable cloud instance with processor Intel(R) Xeon(R) E5-2673 v4 running at 2.30GHz and 425GB of RAM. For a fair comparison all algorithms were executed in single thread mode. The speedups given by MASS-UMAP against linear MSE search were computed by pre-loading all the image archive and query images in memory, so the reported results for the linear MSE search are the fastest possible, with no disk access. While this approach is useful for testing purposes, in real applications this is usually not possible because of memory limits or operational choices, and thus the gap between MASS-UMAP and MSE search will skew even more in favor of the former because of less or no needs of disk access. We want to remember that the entire archive and query embeddings can fit in 342598 * 5 * 4 bytes = 6.5 MB of memory, while the original radar scans account for more than 5 GB of data even after the 64 x 64 pixel resize.

\section{Discussion} \label{discussion}

The MASS-UMAP method proved to be a flexible and performant method: not only it allows for searching analog sequences of arbitrary length in a few seconds over several years of data archive, but it also improves accuracy over published results~\cite{pca2015}. While in this work we focused our analysis only on the minimization of MSE as objective, any positive distance measure can be used to tune the dimensionality reduction. An example of this would be using a distance measure that is robust to a certain degree of rotation or translation, like the Complex wavelet structural similarity~\cite{sampat2009complex}, allowing to find analogs accounting for a certain degree of displacements or rotation~\cite{altencia2015}. The MASS-UMAP method allows also the integration of external variables: synoptic or seasonality descriptors can be integrated by concatenating the desired variables to the UMAP embedding generated for each image and weighted during MASS search. Finally, UMAP neighbors parameter $n$ allows to derive the embedding distribution that finetunes the search results for a specific number of top $k$ analogs. 
We believe that our verification setup was extensive enough that our findings about the optimal values of $d$, $n$ and $k$ can be reused as baseline parameters at least for other radar dataset, but we envision the use of the method also for any other remote sensing applications where spatiotemporal search is needed.

The drawback of our solution is that its flexibility comes with a price: some combination of parameters can give worse result than PCA. To avoid this edge cases a proper verification like the setup proposed in this paper is needed. We show UMAP gives substantially better results than PCA on all reasonable number of dimensions in this setup. The same warning holds for MASS: to avoid search results with spurious matches we provide the top-$k$ MSE reordering mechanism to filter spurious matches as analogs.

As future research directions we plan to use this methodology as an operational application in probabilistic precipitation nowcasting. Another possibility we envision the usage of ensembles of UMAP trained with different configurations and metric functions to improve the retrieval of analogs in embedded space.

\section{Conclusions} \label{conclusion}
In this work we presented an approach to reduce the computational complexity of analog search. Instead of computing the MSE between a search query and the whole historical archive set we demonstrated the efficiency of a combined approach based on 3 steps: dimensionality reduction, fast search in constant time in the embedding space with MASS and MSE-based reordering on a subset of potential candidates, reducing the computational burden by a factor of 20. We also assessed the performances of MASS-UMAP when searching for sequences of different length. In addition compared UMAP against PCA, showing that UMAP ca use a much smaller number of components, leading to superior performances and less memory requirements. These results make the MASS-UMAP approach appliable for nowcasting applications, where efficiency in finding analogs can lead to more accurate and precise predictions in very short time windows.

\vspace{6pt} 

\authorcontributions{conceptualization, G.F.; methodology, G.F. and G.J; software, G.F.; validation, G.F. and L.C.; formal analysis, G.F.; investigation, G.F.; resources, C.F.; data curation, M.P.; writing--original draft preparation, G.F., L.C. and G.J.; writing--review and editing, C.F.; visualization, L.C.; supervision, C.F..; project administration, C.F.; funding acquisition, C.F.}

\funding{Computing resources partially funded by the Microsoft Azure Grant AI for Earth "Modeling crop-specific impact of heat waves by deep learning" assigned to C.F.}


\conflictsofinterest{The authors declare no conflict of interest.} 

\abbreviations{The following abbreviations are used in this manuscript:\\

\noindent 
\begin{tabular}{@{}ll}
MSE & Mean Squared Error\\
PCA & Principal Component Analysis\\
UMAP & Uniform Manifold Approximation and Projection\\
MASS & Mueen's Algorithm for Similarity Search\\
AnEn & Analog Ensemble
\end{tabular}}

\appendixtitles{no} 
\appendix
\section{}
\unskip
\subsection{} \label{extrafigures}

\begin{figure}[H]
    \centering
    \includegraphics[height=3.3 cm]{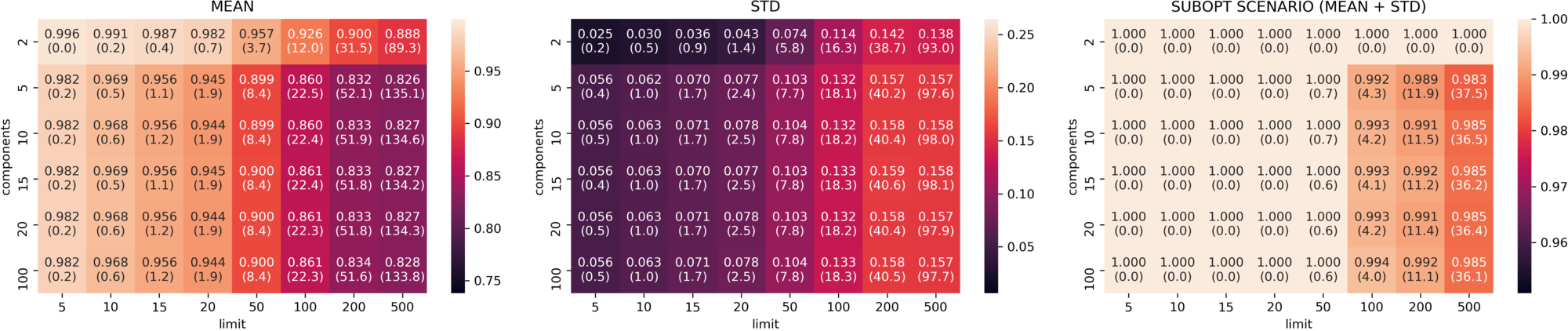}
    \caption{Jaccard results for UMAP models trained with neighbors $n = 100$.}
    \label{fig:umapj100}
\end{figure}

\begin{figure}[H]
    \centering
    \includegraphics[height=3.3 cm]{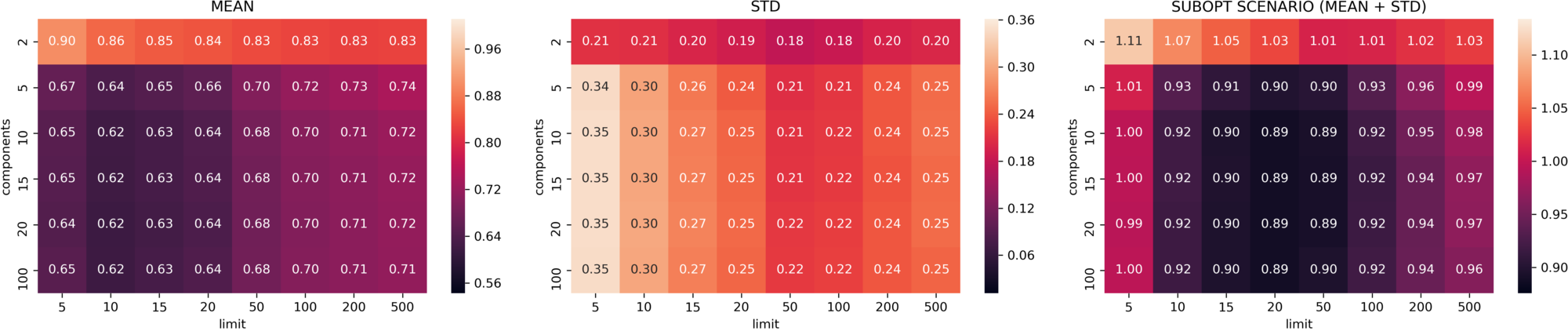}
    \caption{Canberra results for UMAP models trained with neighbors $n = 5$.}
    \label{fig:umapcan5}
\end{figure}

\begin{figure}[H]
    \centering
    \includegraphics[height=3.3 cm]{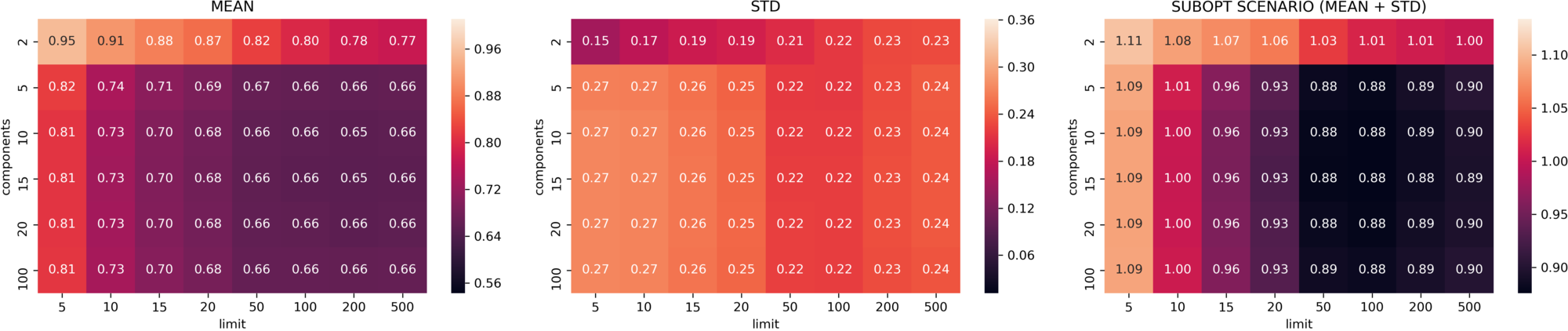}
    \caption{Canberra results for UMAP models trained with neighbors $n = 10$.}
    \label{fig:umapcan10}
\end{figure}

\begin{figure}[H]
    \centering
    \includegraphics[height=3.3 cm]{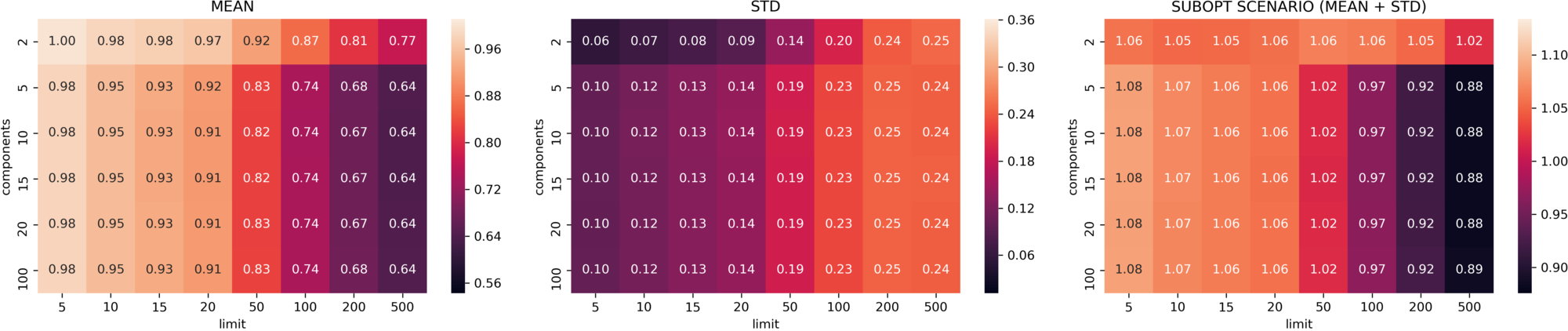}
    \caption{Canberra results for UMAP models trained with neighbors $n = 50$.}
    \label{fig:umapcan50}
\end{figure}

\begin{figure}[H]
    \centering
    \includegraphics[height=3.3 cm]{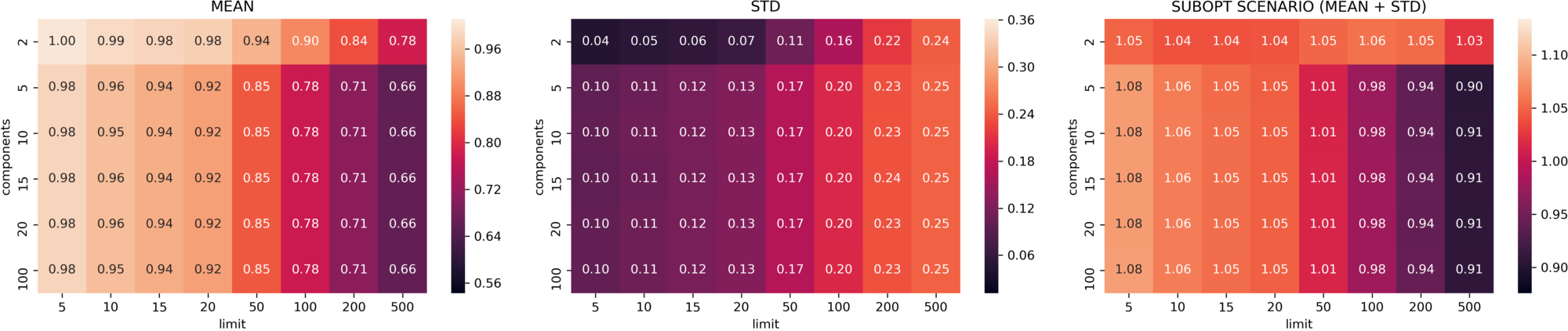}
    \caption{Canberra results for UMAP models trained with neighbors $n = 100$.}
    \label{fig:umapcan100}
\end{figure}

\begin{figure}[H]
    \centering
    \includegraphics[height=3.3 cm]{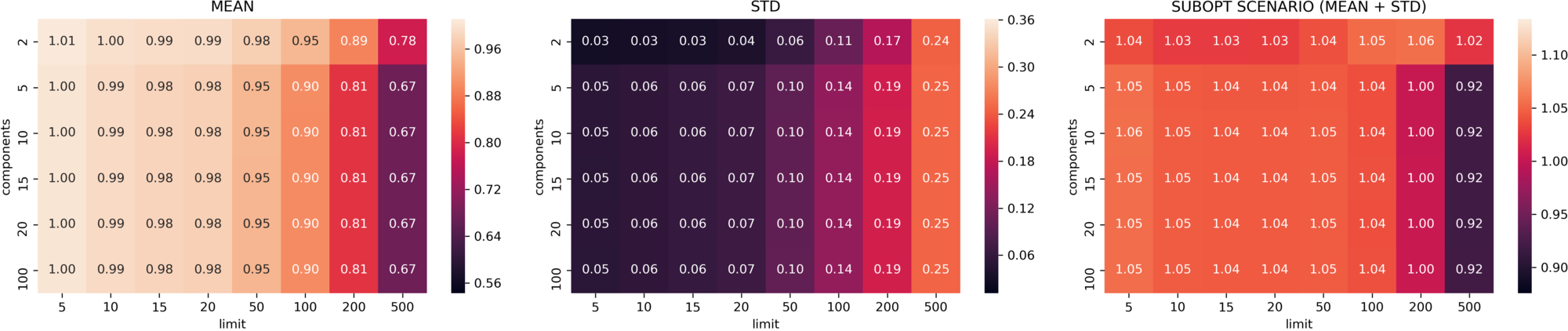}
    \caption{Canberra results for UMAP models trained with neighbors $n = 200$.}
    \label{fig:umapcan200}
\end{figure}

\begin{figure}[H]
    \centering
    \includegraphics[height=3.3 cm]{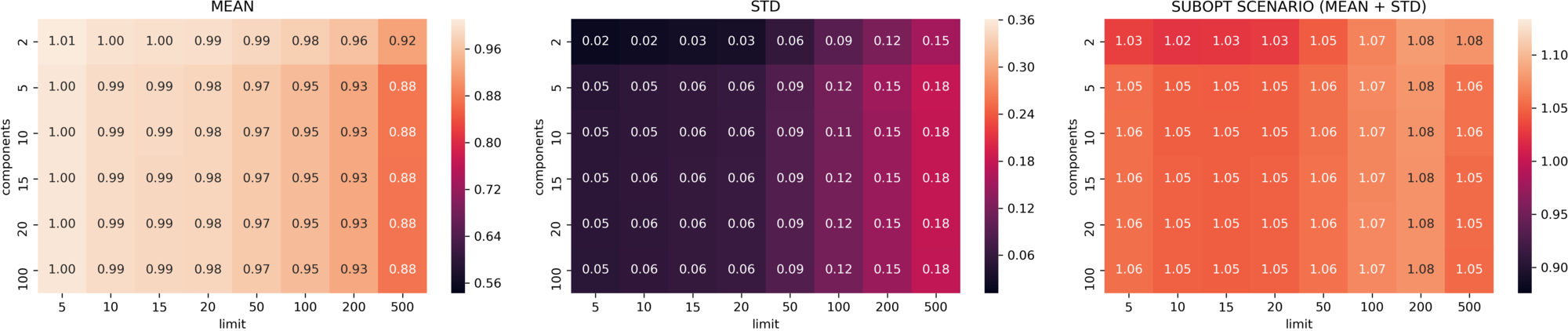}
    \caption{Canberra results for UMAP models trained with neighbors $n = 1000$.}
    \label{fig:umapcan1000}
\end{figure}

\reftitle{References}
\externalbibliography{yes}
\bibliography{main}

\end{document}